\documentclass[lettersize,journal]{IEEEtran}
\usepackage{amsmath,amsfonts}
\usepackage{algorithmic}
\usepackage{array}
\usepackage{textcomp}
\usepackage{stfloats}
\usepackage{url}
\usepackage{verbatim}
\usepackage{graphicx}
\usepackage{cite}
\usepackage{booktabs, tabularx}
\usepackage[referable]{threeparttablex}
\usepackage[ruled,vlined,linesnumbered ]{algorithm2e}
\usepackage{subcaption}

\begin{document}

\title{Towards Connected Smart Work Zones: Advancing Work Zone Management through Improved Connectivity}

\author{Mariam Nour, Mohamed H. Zaki ~\IEEEmembership{Member,~IEEE}, and Mohamed Abdel-Aty~\IEEEmembership{Senior Member,~IEEE}
       
\thanks{Mariam Nour and Mohamed Abdel-Aty are with the Department of Civil, Environmental and Construction Engineering, University of Central Florida, Orlando, FLorida, USA (emails: mariam.nour@ucf.edu; m.aty@ucf.edu)}

\thanks{Mohamed H. Zaki is with the Department of Civil and Environmental Engineering, Western University, London, Canada (email:mzaki9@uwo.ca)}}



\maketitle

\begin{abstract}
Work zones play a key role in road and highway maintenance but can lead to significant risks to both drivers and workers. Smart Work Zones (SWZs) have emerged as a potential solution, offering decision-makers real-time insights into the status of the work zone. By utilizing work zone barrels equipped with sensors and communication nodes, SWZs facilitate collecting and transmitting critical data, including location, traffic density, flow patterns, and worker proximity alerts.
In collaboration with the Florida Department of Transportation (FDOT), this study addresses work zone barrel connectivity requirements while considering a cost-effective, low-power, and low-maintenance solution. While the broader project aimed to create a complete SWZ system for the localization of work zone barrels, this paper proposes a novel relay node selection algorithm integrated with Bluetooth Low Energy (BLE) technology to enhance network performance. The proposed algorithm enhances the communication network performance by selecting specific nodes as relay points, avoiding message flooding in the network. It demonstrates an improvement in message delivery rates, achieving up to a 40\% increase over existing methods while ensuring balanced load distribution among nodes. Moreover, it maintains an 80\% message delivery rate while minimizing power consumption, outperforming other approaches.
This improvement in communication efficiency is critical, as it ensures the accurate transmission and delivery of vital work zone data, allowing for faster and more informed decisions to enhance work zone safety and management.
\end{abstract}


\section{Introduction}
Work zones, defined as areas of the road under construction or maintenance, present unique challenges where the safety of drivers is greatly affected by disruptions to normal traffic operations. These challenges can include: sudden lane reductions due to closures, reduced visibility, and the presence of construction workers and equipment. 
Hence, Smart Connected Work Zones (SWZs) have emerged as a promising solution for enhancing both efficiency and safety within work zones~\cite{nnaji2020improving}. By equipping work zone barrels and equipment with sensors and communication devices, SWZs facilitate the collection and transmission of real-time data, such as barrel location, traffic density, flow patterns, and worker proximity alerts~\cite{park2017improving}. SWZs have the potential to enhance overall safety within the work zone by providing critical information that can be used to address potential hazards ~\cite{wang2013evaluation,awolusi2019active}.
For example, if barrels that guide traffic during lane closures are displaced from their original locations, it can cause confusion, especially in low-visibility conditions, and lead to potential crashes. SWZs offer the capability to detect and communicate the position of these barrels, allowing decision-makers to react quickly and prevent accidents.
Furthermore, the connectivity within SWZs aligns with the broader vision of integrating the Internet of Things (IoT) within smart cities, where all elements are interconnected. This integration paves the way for SWZs to connect seamlessly with other intelligent transportation systems, connected vehicles, and city-wide networks~\cite{knickerbocker2021deploying}.
Apart from the potential advantages of SWZs, there are several challenges. Establishing an efficient communication strategy for SWZs is a key factor, where messages must be delivered to the final destination in a reliable and efficient manner.
These communication solutions should be cost-effective to prevent a significant increase in deployment costs, ensuring accessibility and scalability for different layouts~\cite{vorhes2022low}.  Additionally, the communication solution must be low-power and low-maintenance, as it is impractical to replace batteries for a large number of deployed sensors. Therefore, the sensor nodes attached to the barrels must operate with minimal energy consumption while ensuring reliable transmission of the collected data \cite{nemade2020iot}. Moreover, minimizing maintenance requirements is essential to reduce operational disruptions and ensure continuous functionality in dynamic work zone environments.

On the other hand, work zone barrel connectivity has unique features that distinguish it from traditional fixed wireless sensor networks (WSNs) or highly dynamic vehicular ad hoc networks (VANETs). Work zone barrels are not stationary and can be moved around, but they are also not expected to be highly dynamic since they are intended to be placed in specific predefined locations. This characteristic of connected smart barrels requires flexible networking capabilities. The challenge lies in establishing a cohesive and reliable network for these barrels, which operate in constantly changing environments. 
Also, the unique placement of work zone barrels in a semi-static linear configuration, which is intended to limit access to work zones, can present difficulties in establishing efficient connections between them \cite{jawhar2011linear}. This becomes particularly apparent when the barrels are located beyond the range of the central node responsible for collecting and processing all data. Overcoming these challenges requires innovative communication solutions that can adjust to the distinct characteristics of work zone barrels while establishing a strong and effective network infrastructure to meet their specific needs.

In collaboration with the Florida Department of Transportation (FDOT), this study focuses on meeting work zone barrel connectivity requirements by developing a cost-effective, low-power, and low-maintenance solution. While the broader project aimed to create a complete SWZ framework capable of localizing work zone barrels under these constraints, this paper focuses on enhancing the communication network performance by proposing a novel relay node selection algorithm integrated with Bluetooth Low Energy (BLE) technology. 
The proposed algorithm aims to enhance the efficiency and dependability of the work zone network, enabling effective and low-power sharing of crucial work zone information. To evaluate its performance, the proposed algorithm is compared against three other relay approaches: a basic BLE method where all nodes function as relays, a method that randomly selects relay nodes, and a k-nearest neighbor-based clustering algorithm.
Hence, the contributions of this study are as follows:

\begin{enumerate}
 \item Defining the unique requirements and features of a Smart Work Zone network.
    \item Proposing a clustering-based relay node selection algorithm to overcome the challenges of a linear topology for a SWZ network in a work zone environment.
    \item A comparative analysis that evaluates the proposed algorithm's performance against three other relay node selection algorithms.
    
\end{enumerate}

The rest of the paper is organized as follows: Section 2 presents a review of related work in Smart Work Zones, Linear Wireless Sensor Networks, and BLE and identifies the gaps in the literature. Section 3 details the proposed methodology and the performance analysis metrics. Section 4 outlines the simulation setup and the results of these simulations. Finally, the paper concludes with a summary of the work, followed by a discussion of its implications and potential directions for future research.


\section{Literature Review}
\subsection{Connectivity in Smart Work Zones}

Implementing connected SWZs can play a crucial role in collecting and providing real-time information about closures or crashes in those areas \cite{li2018estimating}. This encourages better decision-making, alternative route choices, easing freeway congestion, and ultimately creating a safer and more efficient environment for motorists and workers \cite{pant2017smart}. 
Early studies exploring integration of technology within work zones included the implementation of dynamic lane merging systems \cite{harb2009two,harb2011two}, dynamic variable speed limit strategies \cite{abdel2008dynamic}, queue-warning systems \cite{tudor2003deployment,sullivan2005work}, and dynamic speed display signs \cite{teng2004evaluation}. 
More recent studies implement intrusion prevention alerts \cite{theiss2018closed,abdallah2024advancing}, sensor devices such as LiDARs \cite{amir2021swzlidar},  remote traffic microwave sensors (RTMSs) \cite{li2018estimating} and smart arrow boards \cite{knickerbocker2021deploying}.
Once real-time traffic data is collected, efficient communication networks are needed to realize the full potential of a fully integrated SWZ. By using cellular leased lines, the authors in \cite{lou2005network} were able to connect multiple work zone barrels with a central node, without range restrictions as compared to other wireless sensor network technologies. Similarly,
the work in \cite{sullivan2005work} has implemented smart work zone barrels to establish an adaptive queue-warning system. The smart barrels were equipped with multiple sensors to track vehicle speed within the work zone. However, these approaches present challenges, including the requirement for a clear line of sight to the central node and the ongoing cost of monthly cellular fees.

In a study by Zhang et al.\cite{zhang2012next}, a complete smart barrel system was developed to enhance work zone safety. The system collects traffic data using Doppler radar sensors, which is then transmitted using Zigbee technology. However, the evaluation was limited to a star network configuration, where each node communicated directly with its corresponding radar node. This setup posed a risk of network segregation if any radar node failed. Additionally, the use of Zigbee technology introduced the challenge of high energy consumption due to the necessity of a router.
A smart connected work zone was implemented in \cite{martin2016wsn} using the 868 MHz SRD band for the system communication. Their choice was based on the communication range provided by the band and that it can be used without a license. Since the network was built in a star topology, nodes have to be within the range of the central node for successful transmission of the collected sensor data. 

In contrast to these earlier studies, the work presented in this paper focuses on creating a comprehensive smart work zone barrel network by leveraging cost-effective, low-power BLE sensors. 
Our proposed approach ensures that sensors on all barrels can successfully transmit their data, regardless of their physical location or distance from that central node by choosing the most appropriate relay nodes.

\subsection{Linear Wireless Sensor Networks}

For the effective deployment of a connected SWZ, it's crucial to consider a suitable communication network topology as it determines how data is transmitted between the nodes within the work zone. 
In this context, the Linear Wireless Sensor Network (LWSN) emerges as the preferred network topology for SWZs because it aligns well with the specific linear structure of work zones. LWSN is a type of Wireless Sensor Network (WSN), a broader category of networks that have been widely used in various transportation-related applications due to their ability to collect and communicate data among nodes without the need for physical connections \cite{bernas2018survey}. 
WSNs employ various network topologies such as clusters, linear, trees, grids, and meshes, depending on the application requirements\cite{alam2020internet}. Table (\ref{tab:nettop}), shows the advantages and limitations of some of these topologies.

\begin{table*}[!ht]
  \centering
    \begin{threeparttable}
    \caption{Comparing different network topologies}
    \label{tab:nettop}
    \begin{tabular}{>{\centering\arraybackslash}p{2cm} >{\centering\arraybackslash}p{5cm} >{\centering\arraybackslash}p{5cm}}
    \toprule 
            \textbf{Network Topology} & \textbf{Advantages} & \textbf{Limitations}\\

    \midrule 
\textbf{Tree}  
               & \begin{minipage}[t]{4cm}
        \begin{itemize}
          \item Easy to expand
          \item Simple management
        \end{itemize}
      \end{minipage} 
      &\begin{minipage}[t]{4cm}
        \begin{itemize}
          \item Vulnerable to node failure
          \item Performance drops with more nodes
        \end{itemize}
      \end{minipage}\\  \midrule 
\textbf{Mesh}     &
             \begin{minipage}[t]{4cm}
            \begin{itemize}
                \item Redundant paths
                \item No single failure point
            \end{itemize}
            \end{minipage}   &     
            \begin{minipage}[t]{4cm}
            \begin{itemize}
                \item Complex setup
                \item Requires routing at each node
            \end{itemize}
            \end{minipage}                       \\  \midrule

\textbf{Linear}   &
\begin{minipage}[t]{4cm}
\begin{itemize}
   \item Simple setup
    \item Cost-effective for small networks
\end{itemize}
\end{minipage}  & \begin{minipage}[t]{4cm}
\begin{itemize}
     \item Prone to node failure
    \item Higher power use with distance
    \item Increased delays in larger networks
\end{itemize}
\end{minipage}   \\ 
    \bottomrule 
\end{tabular}
    \end{threeparttable}  
\end{table*}

LWSNs are a special class of WSNs characterized by nodes arranged in a strictly linear or semi-linear configuration \cite{jawhar2011linear}.   
They are used in various applications in monitoring and sensing environments, such as power transmission lines monitoring \cite{kong2021adaptive}, underwater sensors \cite{fattah2020survey}, and pipeline infrastructure monitoring \cite{jawhar2007framework}. 
This configuration is well-suited to the linear layout of most work zones, making LWSNs ideal for establishing efficient communication within a SWZ.
However, LWSN implementation poses various technical challenges \cite{chen2008chain}. In most LWSN configurations, a central node functions as the data sink, collecting and processing information from other nodes in the network. If a linear network chain becomes longer, some nodes may fall outside the communication range of the central node, which would disrupt the network operation \cite{zhang2011analysis}. To resolve this issue, nodes closer to the sink would not only collect and transmit their own data but also relay data from out-of-range nodes. As a result, one of the key challenges to implement such LWSNs would be energy efficiency, as the nodes closest to the sink tend to have a higher load due to their role in forwarding data from multiple out-of-range nodes \cite{carsancakli2022reliability}.
To overcome this challenge, the communication approach used should integrate an efficient relaying mechanism. This ensures that nodes beyond the reach of the sink node can have their messages successfully received at the sink while considering the energy efficiency aspect \cite{jawhar2009efficient}.
Determining the optimal method for selecting the optimal relay nodes is important while focusing on factors such as fairness and energy efficiency within the communication network \cite{hansen2018relay}.
Therefore, it is critical to adopt efficient data transmission techniques that are specifically designed to address the unique challenges of implementing LWSN within a SWZ \cite{liang2019relay}. Our work introduces a clustering-based relay node selection algorithm for LWSNs within a SWZ. It aims to ensure high message delivery ratios, fairness among nodes regardless of their relative location to the sink, and address power consumption aspects.

\subsection{Bluetooth Low Energy and Relay Node Selection}
To implement a cost-efficient and low-power solution for work zone barrel connectivity, a suitable communication technology needs to be chosen. BLE satisfies these conditions and was chosen as the communication technology in our proposed system. BLE-powered devices can be purchased off-the-shelf, easily configured and are cheap.
In 2010, the introduction of BLE marked a significant leap forward in wireless network technologies, particularly in smartphones, tablets, and across different Internet of Things (IoT) categories such as smart city applications \cite{hasan2022someone}, localization \cite{zhuang2022bluetooth}, health and fitness, and wearables \cite{woolley2017bluetooth}. 
BLE's low-power and long-range features make it a suitable choice for various monitoring and sensing applications, especially in the realm of Internet of Things (IoT). Traditionally, BLE supported only a star topology, where nodes communicate exclusively with a central node \cite{darroudi2017bluetooth}. To overcome this limitation, BLE Mesh was introduced, enabling a many-to-many communication among multiple devices \cite{brandao2020energy}. This allows nodes to operate without a central node and to interact directly with one another \cite{baert2018bluetooth}.
The current implementation of BLE mesh relies on a managed flooding approach to disseminate data within the network \cite{rondon2019understanding}. Flooding is a technique where each node in the network forwards incoming data to all neighboring nodes. This technique ensures widespread dissemination of data, which creates a resilient network against path failures without the need for complicated and power-consuming routing algorithms \cite{de2020experimental}.
Although managed flooding can be suitable for low-energy purposes in some cases, it can also be inefficient in others. Broadcast storms, network delays, packet collisions, and high energy consumption are problems that arise with the flooding approach \cite{wang2022ace}. A possible solution to these challenges would be implementing a routing algorithm, where a specific path for each packet from source to destination is defined \cite{guo2015demand}.
However, LWSNs have unique properties that make adopting widely used routing protocols challenging, especially a network that would be used for work zone barrel connectivity. One of the primary challenges is energy efficiency, as nodes need to consume more energy for route computation and require additional memory to store potential paths. \cite{kong2021adaptive}.
In addition to managed flooding, BLE mesh includes a relaying feature that can further extend the network connectivity. Relaying allows message re-transmissions to reach the destination node when it is out of range of the sending node. Hence, choosing which nodes would implement the relay feature is an important factor to consider while building an efficient network \cite{hansen2018relay}. 
Since the relaying feature can be power intensive, it's crucial to find a balance between ensuring that the network is connected while lowering the energy cost.
Given that the BLE Mesh specifications don't define a specific protocol for relay node selection and leaves it up to the user, in this work, we propose a relay node selection strategy for SWZ barrel connectivity. Our developed algorithm uses a clustering approach, particularly in the context of LWSNs, that can be used in connecting work zone barrels efficiently.
Relay node selection has been widely studied in the field of WSNs \cite{liang2019relay,shukla2020effective} and also specifically in BLE Mesh \cite{hansen2018relay,reno2020relay}. However, the simulation studies carried out in these studies focused on different network topologies other than LWSNs. The application of relay node selection for semi-static networks such as those for SWZs have also not been explored.
In comparison, our proposed approach is unique in that it introduces a novel relay node selection scheme tailored for Linear Wireless Sensor Networks in Smart Work Zones, utilizing BLE mesh to enhance connectivity and performance.


\section{ Clustering-based Relay Node Selection in a BLE Linear Network}

\begin{figure*}[!h]
\begin{center}
\centerline{\includegraphics[width=0.7\textwidth,height=0.32\textwidth]{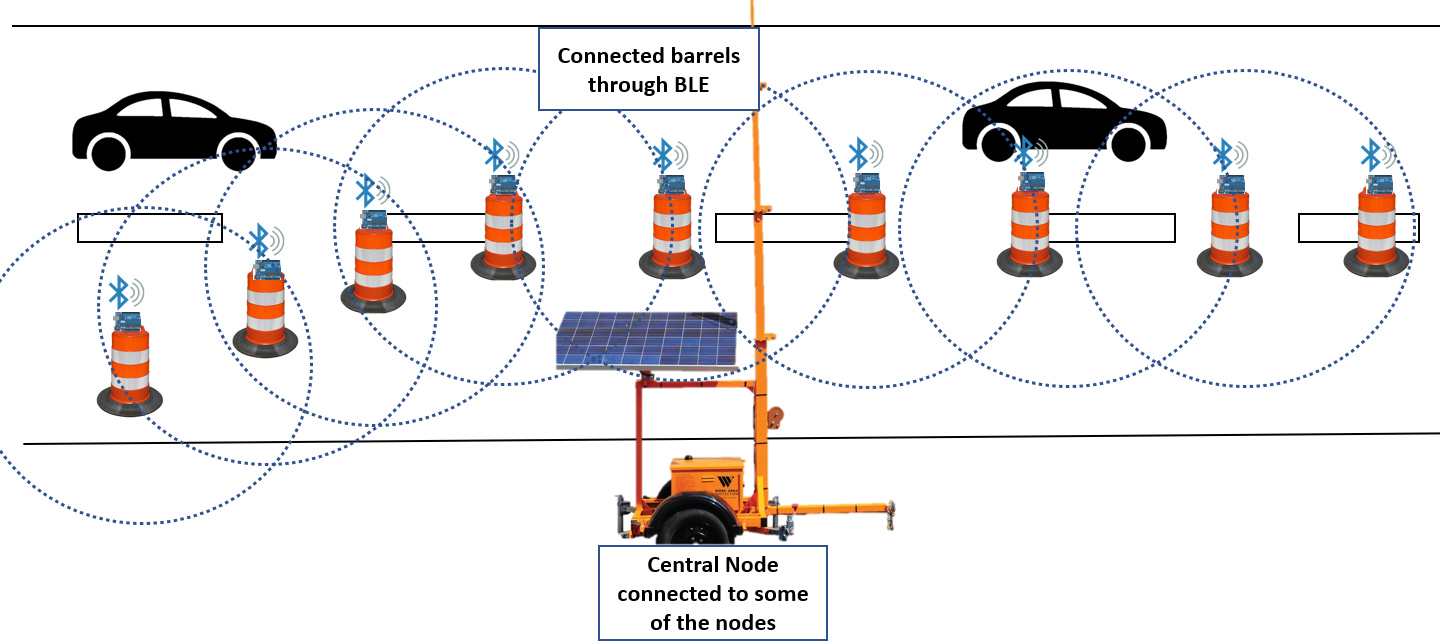}}
\caption{System Framework}
\label{fig:systemarch}
\end{center}
\end{figure*}

Figure (\ref{fig:systemarch}) shows the overall system framework. The SWZ barrels are arranged linearly, each equipped with BLE sensors and various data collection sensors. In line with the FDOT requirements for our collaborative project, GPS sensors were used for location tracking, and accelerometers were included to detect barrel movement or impacts. These sensors are crucial for monitoring the precise position and status of each barrel within the work zone. Additionally, the barrels can be equipped with other types of sensors, such as those for intrusion detection, vehicle counting, or traffic flow monitoring.
A central node along the work zone serves as the sink, receiving all collected data. As the number of barrels increases, cost-efficiency becomes essential, and frequent battery replacements are impractical, necessitating low-power strategies to extend battery life. Hence, BLE was chosen for its ability to balance these constraints. The semi-static nature of work zones, where barrels may be moved or impacted, required a network capable of efficiently broadcasting the precise location and status of each barrel without continuous transmission, which would drain power. Therefore, a possible strategy can be that nodes only transmit when necessary, such as when a barrel moves or detects a significant event, ensuring reliable data delivery to the central node while optimizing power consumption.

A clustering-based relay node selection (C-RNS) algorithm is proposed to enhance node communication efficiency. The objective of the proposed C-RNS algorithm is to effectively distribute packets within the network to reach the sink, all while maintaining low power consumption and fairness amongst the relay nodes. The proposed algorithm consists of two phases: the first phase involves selecting the relay nodes, followed by the data dissemination phase.

\subsection{Relay Node Selection Process}

The proposed algorithm depends on each node outside the sink's range to choose a relay node based on an assigned score. The score is computed using the total number of neighbors, the distance from a specific node, and the total number of nodes choosing this neighbor as a relay.

Initially, each node's initial score ($Score_{initial}(j)$) is determined by the count of its in-range neighbors represented by equation (\ref{eq:scoreinit}). From each node’s perspective, a neighbor vector is then created with the initial scores of its neighboring nodes. The score calculation process begins at each node in ascending order of IDs, facilitating the selection of relay nodes.

\begin{equation}
\label{eq:scoreinit}
    Score_{initial}(j) = NeighborVector(j)
\end{equation}

For a given selecting node $i$, the neighbor nodes are sorted in descending order, and the distance from the selecting node is subtracted from their initial scores, represented by equation (\ref{eq:scoreati}). This adjustment ensures that, among neighbor nodes with similar initial scores, the closer neighbor node to the selecting node is given a higher score. 

\begin{equation}
\label{eq:scoreati}
    Score_{i}(j) = Score_{initial}(j) - \frac{distance(i,j)}{R}
\end{equation}

The neighbor with the highest score is then chosen as a relay node for the selecting node $i$, and this neighbor's initial score is reduced by 1 (equation \ref{eq:scorefinal}). This reflects that this neighbor is now being utilized as a relay and has a decreased capacity to serve other nodes. In scenarios where neighboring nodes have equivalent scores, the node with the lowest ID is selected.

\begin{equation}
\label{eq:scorefinal}
    Score_{initial}(j) = Score_{initial}(j) - 1
\end{equation}

These steps guarantee a balanced distribution of the relaying load, considering both neighborhood count and proximity to nodes in need of relay services. As every node undergoes this selection process, the network becomes fully connected, with nodes either serving as direct relays or being indirectly connected via a series of relay nodes. After completing the relay node selection phase, the network forms clusters of relay nodes and the nodes they serve directly. This structure forms the foundation for efficient data routing towards the sink.


Algorithm (\ref{alg:main1}) presents the pseudocode for the relay node selection algorithm (C-RNS).
\begin{algorithm}[!h]
\SetAlgoLined
\KwData{Nodes positions $(X,Y)$, Communication range $R$}
\KwResult{List of relay nodes}
Initialization: $NeighborMatrix = 0$, $relayNodes = 0$, $NeighborVector = 0$, $Score = 0$\;
\For{each node $i$}{
    \For{each node $j$}{
        \If{$distance(i,j) < R \&\& distance(i,j) > 0$}{
            $NeighborMatrix(i,j) = 1$\;
            $NeighborVector(j) = \sum NeighborMatrix(j,:)$\;
            $Score(j) = NeighborVector(j)$\;
        }
    }
}
\For{each node $i$}{
    \If{$distance(i, \text{sink}) \geq R$}{
        $potentialRelays = \text{find}(NeighborMatrix(i,:))$\;
        \For{each relay $r$ in $potentialRelays$}{
            $Score(r) = Score(r) - \frac{distance(i,r)}{R}$\;
        }
        $chosenRelay = \text{argmax}(Score(potentialRelays))$\;
        \If{$chosenRelay$ exists}{
            $relayNodes(chosenRelay) = 1$\;
            $Score(chosenRelay) = Score(chosenRelay) - 1$\;
        }
    }
}
\caption{Clustering-based relay node selection algorithm with Euclidean distance}
\label{alg:main1}
\end{algorithm}
The C-RNS algorithm begins by iterating through each node in the network, calculating their initial score based on the count of in-range neighbors and creating a neighbor matrix with the scores of all other nodes (lines 3-7). For each node, the algorithm sorts its neighbors in descending order based on their scores and adjusts these scores by subtracting the distance from the selecting node (lines 12-16). The algorithm then selects the neighbor with the highest adjusted score as the relay node for the current node, updating the relay Nodes array and decrementing the chosen relay node's score (lines 17-20). 
As each node independently undergoes this selection process, the resulting network achieves full connectivity, with nodes serving as either direct relays or indirectly connected through a series of relay nodes.

\subsection{Data Dissemination Process} 

\begin{figure*}[!h]
\begin{center}
\centerline{\includegraphics[width=0.55\textwidth,height=0.3\textwidth]{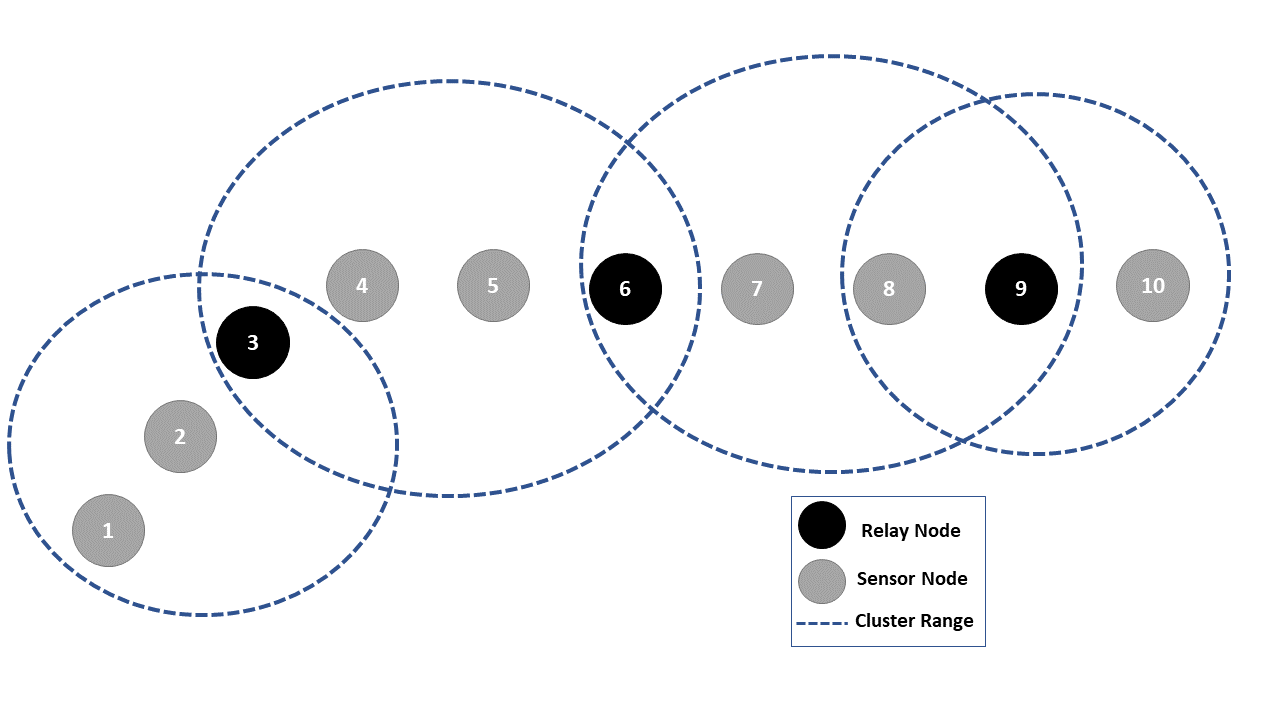}}
\caption{Resulting network after applying C-RNS}
\label{fig:clustering}
\end{center}
\end{figure*}

Figure (\ref{fig:clustering}) illustrates an example of a fully connected network post the execution of the C-RNS algorithm. The figure highlights the distinct clusters formed, where the relay node can be regarded as the cluster head. 

Now that all nodes are connected data dissemination can begin. Since this is a linear topology where the sink node can be placed anywhere along the length of the network, the nodes near and in range of the sink are inherently at an advantage for a higher packet delivery ratio. Therefore, to give all other nodes' messages a fair opportunity to reach the destination, nodes further away from the sink have a higher network transmission frequency. Hence, the farther nodes send more network-layer messages for a greater chance of reaching the sink node. Another thing to note is that the nodes that are in the range of the sink don't participate in the relay node selection algorithm as they don't need to choose a relay node.

\subsection{Performance Evaluation Process}

In order to assess the effectiveness of the relay node selection algorithm proposed, several performance evaluation metrics are introduced and employed for the analysis and comparison of various relay node selection algorithms. These metrics include packet delivery ratio, relay load balancing and power consumption.


\subsubsection{Packet Delivery Ratio}

Network-wide packet-delivery-ratio (PDR) was calculated for the various scenarios using equation (\ref{eq:pdr_net}). This metric provides an understanding of overall network efficiency and the percentage of messages received by the sink node. The higher the PDR value, the better the network is performing.

\begin{equation}
\label{eq:pdr_net}
    PDR_{network} = \frac{\sum_{n=1}^{N1}R_n }{\sum_{n=1}^{N2} S_{n}}
\end{equation}

Where $N1$ is the total number of destination nodes
, $N2$ is the total number of source nodes
, $R_{n}$ is the total number of packets received by all destination nodes
and $S_n$ is the total number of packets sent by all source nodes.

Although the aggregate Packet Delivery Ratio (PDR) is useful for evaluating the performance of the network as a whole, it does not reveal the performance at each individual node. Evaluating the PDR at a per-node level can provide insight into the fairness aspect, as it offers an understanding of the effectiveness of the relay node selection algorithm in its ability to successfully transmit messages from all nodes to the sink, specifically those that are out of range of the sink node. Hence, the per-node PDR is calculated for each node according to equation (\ref{eq:pdr_node}).

\begin{equation}
\label{eq:pdr_node}
    PDR_{n} = \frac{P_{received} *100}{P_{sent}}
\end{equation}

Where $P_{received}$: is the total number of messages received by the sink from node n and
$P_{sent}$: is the total number of messages transmitted by node n.

\subsubsection{Relay Load Balancing}

Ensuring load balancing among relay nodes is a crucial metric to consider when implementing a relay node selection algorithm \cite{zhang2020cooperative}. The traffic load, represented by the number of relayed messages in the network layer for each relay node, provides insights into the fairness of the algorithm in distributing the relaying load. Effective load balancing is essential as it impacts several aspects of the network's performance. Firstly, it enhances the overall efficiency by preventing congestion in specific relay nodes, ensuring a more balanced distribution of the workload \cite{Jiang2011load}. Additionally, balanced relay loads contribute to energy efficiency, as it mitigates the risk of certain nodes depleting their power resources faster than others \cite{Kong2016relaychain}. This consideration becomes particularly significant in maintaining the longevity and reliability of the network. 

\subsubsection{Power Consumption}

Analyzing the power consumption of relay nodes is essential for understanding the overall lifespan of the network \cite{brandao2020energy}. This is because relay nodes are vital in keeping the network connected.
During operation, nodes go through multiple phases: idle, transmitting, listening and sleeping. Each phase has a different energy consumption profile determined by hardware-dependant parameters. Therefore, the average power consumption of each relay node can be computed and averaged over the simulation time using equation (\ref{eq:energy}).
\begin{equation}
\label{eq:energy}
    C_n = T_{tx}i_{tx} + T_{listen}i_{listen} + T_{sleep}i_{sleep}
\end{equation}

Where $C_n$ is the average current consumption for node $n$, $T_{tx}$ is the average transmission time, $i_{tx} $: Average transmission current, $T_{listen}$ is the average listening time, $i_{listen}$: is the current consumed during the listen state, $T_{sleep}$ is the average sleep time and $i_{sleep}$ is the current consumed during the sleep state.


\section{Evaluation}
The design of the proposed approach was a part of our prior work with the Florida Department of Transportation on smart work zones, where we focused on creating a cost-effective, low-energy solution for connecting work zone barrels. The main objective of that project was to develop a network that could accurately localize individual work zone barrels. Building on this foundational work, we extended our research to explore the enhancement of communication strategies within a SWZ network.

In this section, the simulation experiments and their outcomes are presented and designed to assess the effectiveness of the proposed relay node selection algorithm (C-RNS) within a BLE mesh linear network. This evaluation is conducted by comparing three relay node selection strategies. The first approach is where every node functions as a relay node. The second strategy involves the selection of a random number of relay nodes. Lastly, we implement and compare our approach with a relay node selection algorithm that is an updated version of the K-Nearest Neighbor-based (KNN) method, as described in the study by Song et al. \cite{song2017unsupervised}.

\subsection{Simulation Setup}

\begin{figure*}[!ht]
\begin{center}
\centerline{\includegraphics[width=0.8\textwidth,height=0.4\textwidth]{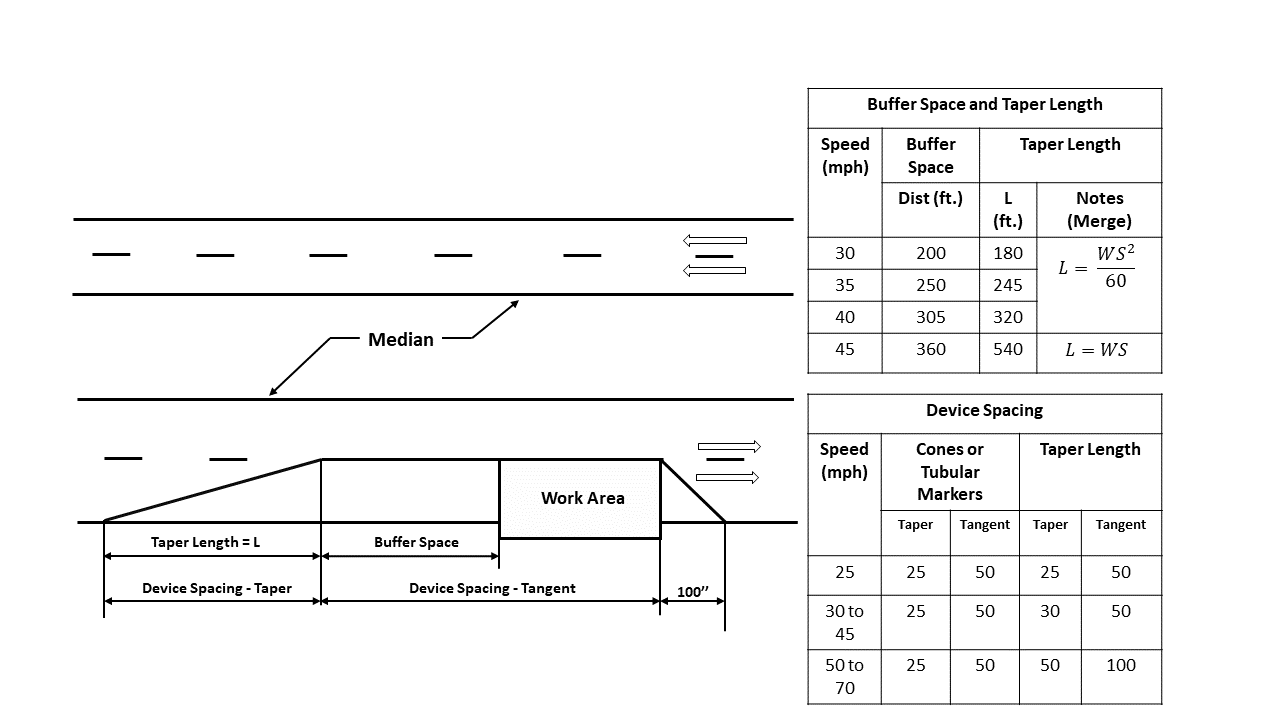}}
\caption{FDoT standards for lane closure setup \cite{fdotwebsite}}
\label{fig:fdot}
\end{center}
\end{figure*}

Following the Florida Department of Transportation Standard Plans of Construction for Lane Closures \cite{fdotwebsite}, the network setup of the work zone barrels was implemented and simulated. The values equivalent to a 45mph freeway (figure \ref{fig:fdot}) were used to calculate the lengths of the taper, buffer, work zone areas, and the spacing between the work zone barrels. 
The overall distance of the closure considered in this study was 1140 feet with a total of 30 work zone barrels and 1 sink node.

The MATLAB Bluetooth toolbox \cite{MathWorksBluetoothToolbox} was then used to simulate the network of work zone barrels fitted with BLE sensor modules. The BLE toolbox enables the generation of realistic Bluetooth network topologies and emulates communication among nodes. Nodes can take different roles based on the BLE mesh specification, such as relay, friend or low-power nodes. Using the work zone barrel network implemented in MATLAB, the relay node selection algorithm based on \ref{alg:main1} was run to decide which nodes would be configured as relays.

The simulation parameters in Table (\ref{tab:params_base}) were used for the four different relay node selection algorithms. The communication range for the base approach where all nodes are relays is 150 meters, while it's 100 meters for the random relays, proposed clustering-based relay node selection algorithm (C-RNS) and the K-nearest Neighbor relay node selection algorithm (K-RNS). This is to further reduce the power consumption of the relay nodes as it's directly proportional to the transmission power. Performance metrics, including packet delivery ratio (PDR), load balancing and energy consumption, were calculated and compared for each simulation scenario.

\begin{table}[!ht]
	\caption{Simulation Parameters}\label{tab:params_base}
	\begin{center}
		\begin{tabular}{l l }
            \toprule
			\textbf{Parameter} & \textbf{Value}  \\\hline

			\textbf{Simulation Time}   & 20 seconds \\
			\textbf{Time-to-live}   & 127 (Default) \\
			\textbf{Transmission rate (Packets/sec)} & 1, 4 \\
			\textbf{Network Size}   & 30 nodes $+$ 1 sink\\
			\textbf{Transmission range} & 100 meters, 150 meters \\
			\textbf{Transmitter power} & 20 dBm \\   
   \hline
		\end{tabular}
	\end{center}
\end{table}


\subsection{Performance Analysis}

To compare the four different relay node selection algorithms, simulations were run to evaluate the performance metrics, including packet delivery ratio (PDR), load balancing and energy consumption. 
Table (\ref{tab:pdr_stats}) presents the overall packet delivery ratio (PDR) for each scenario and the percentage change compared to the base scenario, where all nodes are relays for two different transmission rates. 
Under both transmission rates, C-RNS consistently outperforms the other relay node selection algorithms. For the 1 packet/sec scenario, C-RNS achieves a PDR of $82\%$, showcasing a $6\%$ improvement over the base scenario. Similarly, in the 4 packets/sec scenario, C-RNS scores a PDR of $81\%$, marking a $40\%$ improvement over the base scenario. These results show that C-RNS exhibits consistent and better performance in PDR across different transmission rates, emphasizing its efficacy in improving the reliability of the network.
\begin{table}[!ht]
	\caption{Packet Delivery Ratio}\label{tab:pdr_stats}
	\begin{center}
		\begin{tabular}{l l l l}
  \toprule
			\textbf{Transmission rate}         & \textbf{Scenario}        &\textbf{PDR}          &\textbf{\% change}    \\\hline
			\textbf{1 packet/sec}   &  All relays       & $77\%$     &   base scenario \\
			                        &  Random relays    & $62\%$     &    $-19\%$      \\
			                        &  KNN RNS          & $62\%$     &      $-19\%$   \\
			                        &  C-RNS            & $82\%$     &    $6\%$   \\\hline
                        
			\textbf{4 packets/sec}  &  All relays       &  $58\%$   &    base scenario \\
                                    &  Random relays    &  $66\%$   &     $13.7\%$   \\
                                    &  KNN RNS          &  $63\%$   &    $8.6\%$   \\
                                    &  C-RNS            &   $81\%$  &      $40\%$   \\\hline
  
		\end{tabular}
	\end{center}
\end{table}

\begin{figure}[!h]
\begin{center}
\centerline{\includegraphics[width=0.5\textwidth,height=0.4\textwidth]{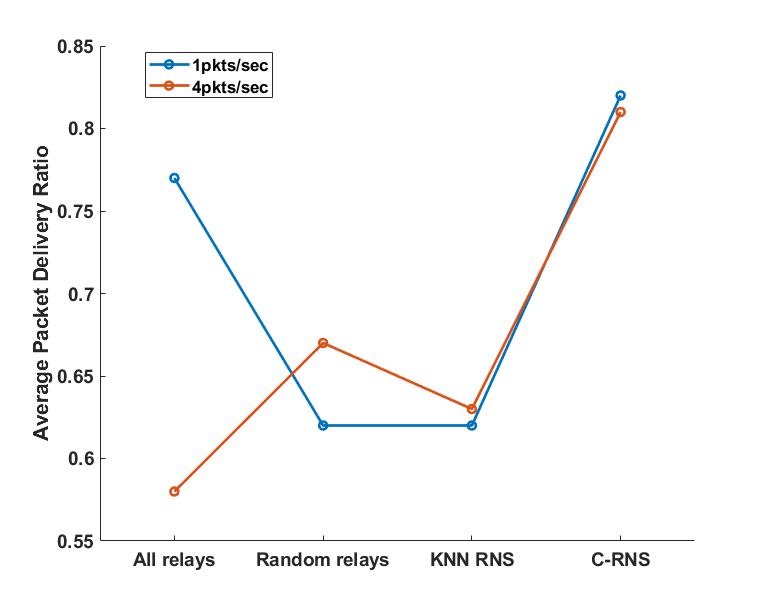}}
\caption{Network-wide PDR for transmission rates: 1pkts/sec and 4pkts/sec}
\label{fig:overallPDR}
\end{center}
\end{figure}

Figure (\ref{fig:overallPDR}) highlights the improvement in the overall network PDR when using C-RNS compared to the other approaches. This boost in efficiency is crucial for ensuring reliable and timely communication within the network, emphasizing the practical significance of employing C-RNS in scenarios demanding enhanced performance and reliability.

\begin{figure}[!ht]
     \centering
     \begin{subfigure}[b]{0.45\textwidth}
         \centering
        \includegraphics[width=\textwidth]{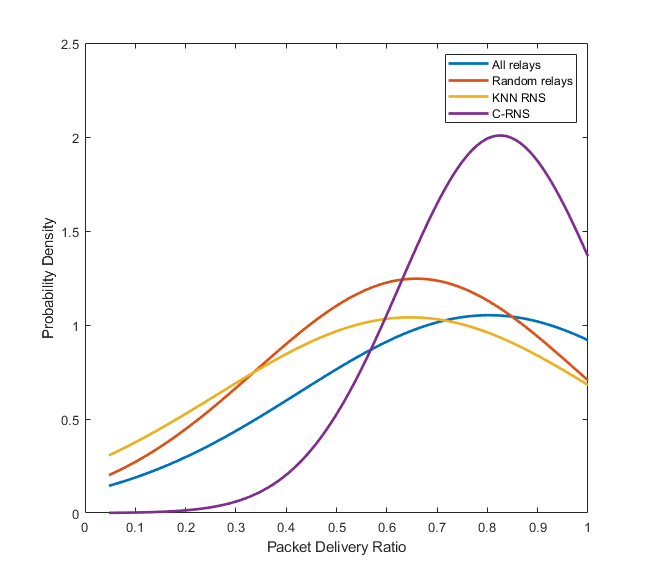}
         \caption{1pkts/sec transmission rate}
         \label{fig:pdr1pkts}
     \end{subfigure}
     \hfill
     \begin{subfigure}[b]{0.5\textwidth}
         \centering
        \includegraphics[width=\textwidth]{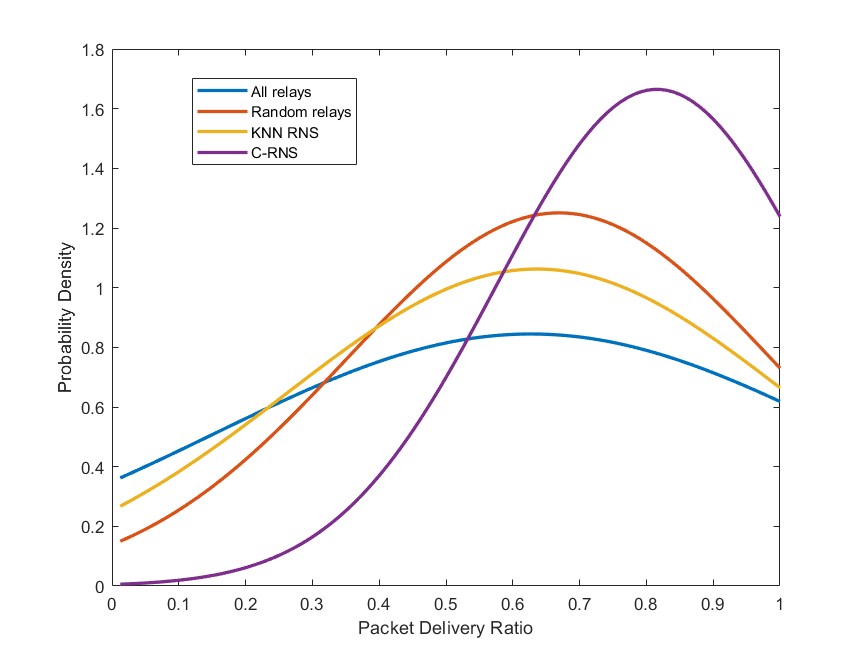}
         \caption{4pkts/sec transmission rate}
         \label{fig:pdr4pkts}
     \end{subfigure}
    \caption{Per Node Packet Delivery Ratio PDF}
        \label{fig:pernodePDR}
\end{figure}

Figures (\ref{fig:pdr1pkts}) and (\ref{fig:pdr4pkts}) illustrate the Packet Delivery Ratio (PDR) density across various relay node selection algorithms at different transmission rates. In both figures, the C-RNS curve reveals that a majority of nodes achieve a PDR exceeding 0.5, with none recording a PDR of 0. This is in contrast to the alternative approaches, where a considerable number of nodes have a PDR of 0. This shows that, for both transmission rates, the proposed C-RNS outperforms the other relay node selection approaches. This can be attributed to the calculations performed within the C-RNS algorithm, which take into account variables such as distance and load balancing. This guarantees that nodes beyond the sink's range have relay nodes within their respective ranges that are not overloaded.

\begin{figure*}[!ht]
    \centering
    \begin{subfigure}[b]{0.3\textwidth} 
        \centering
        \includegraphics[width=\textwidth]{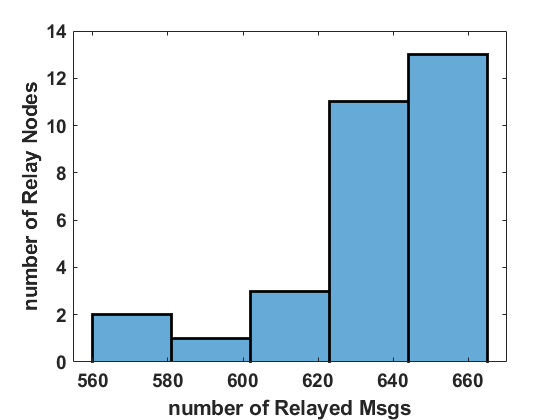}
        \caption{All relays}
        \label{fig:loadAllrelays}
    \end{subfigure}
    \quad 
    \begin{subfigure}[b]{0.3\textwidth} 
        \centering
        \includegraphics[width=\textwidth]{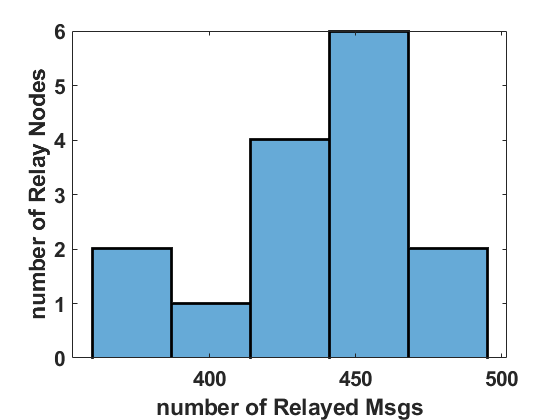}
        \caption{Random}
        \label{fig:loadRand}
    \end{subfigure}
    \begin{subfigure}[b]{0.3\textwidth} 
        \centering
        \includegraphics[width=\textwidth]{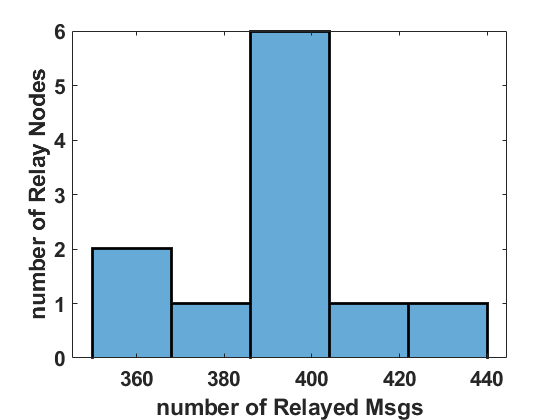}
        \caption{KNN}
        \label{fig:loadKNN}
    \end{subfigure}
    \quad 
    \begin{subfigure}[b]{0.3\textwidth} 
        \centering
        \includegraphics[width=\textwidth]{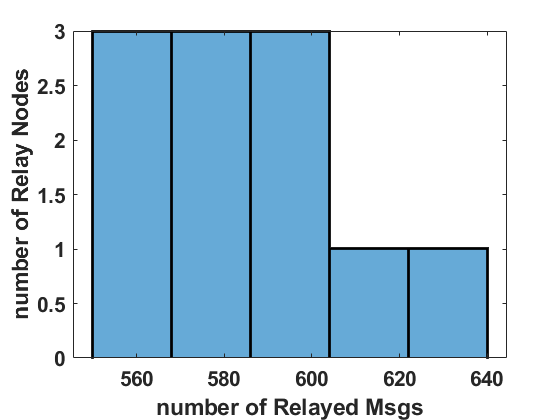}
        \caption{C-RNS}
        \label{fig:loadRNS}
    \end{subfigure}

    \caption{Relay Load Balancing for the different relay node selection approaches}
    \label{fig:loadbalancing}
\end{figure*}

Figure (\ref{fig:loadbalancing}) illustrates the relay load balancing distribution across the four relay node selection approaches. It can be noted that for the C-RNS, most relay nodes are handling a similar number of packets (i.e., most nodes fall into the same or adjacent bins); this indicates a relatively even distribution of load. In contrast to the other relay selection approaches, where the histogram is more spread out with nodes handling significantly different numbers of packets, which indicates an uneven distribution of the relaying load.

\begin{figure}[!h]
\begin{center}
\centerline{\includegraphics[width=0.45\textwidth]{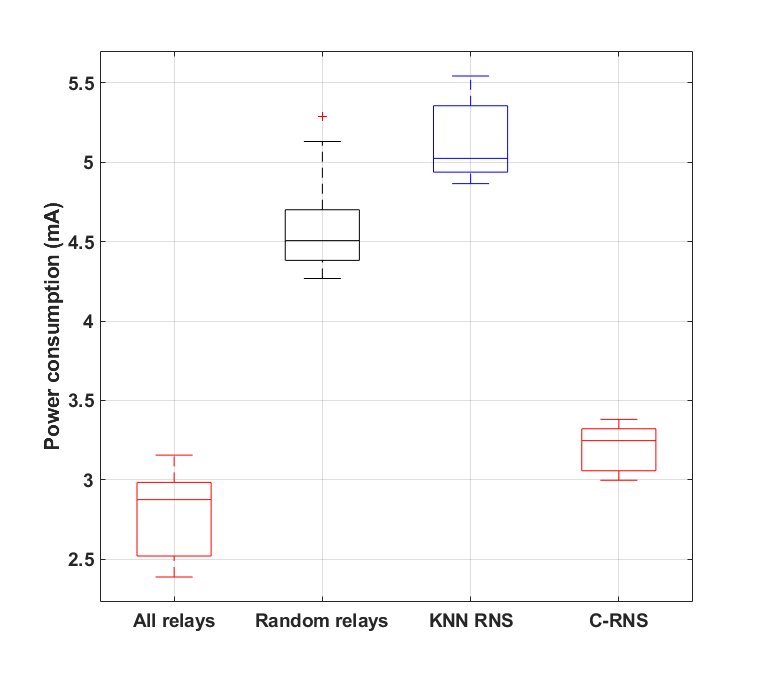}}
\caption{Power consumption (mA)}
\label{fig:current}
\end{center}
\end{figure}

Illustrated in Figure \ref{fig:current} is the average power consumption in mA for the relay nodes for all relay node selection algorithms. Notably, both the all relays and C-RNS approaches show better energy consumption performance when compared to other methods. 
This improvement can be attributed to the effective load balancing aspect, where all relays and C-RNS approaches offer a more efficient distribution of relay load across the nodes in comparison to alternative methods.

\begin{figure}[!h]
\begin{center}
\centerline{\includegraphics[width=0.4\textwidth]{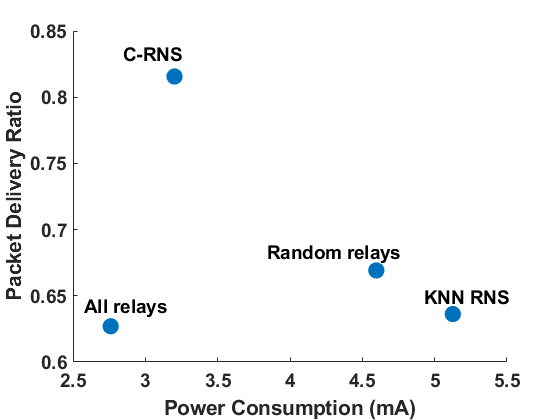}}
\caption{Power consumption (mA) vs PDR}
\label{fig:pdrvspower}
\end{center}
\end{figure}

However, when comparing energy consumption and overall Packet Delivery Ratio (PDR) as shown in Figure \ref{fig:pdrvspower}, C-RNS demonstrates the best performance in the energy consumption vs. PDR trade-off. This suggests a potential slight trade-off between achieving significantly improved PDR and optimizing energy consumption. However, it's important to note here that C-RNS still had the superior per-node PDR, which is a more relevant metric to the work zone barrel application.

\section{Discussion}

This study explores the implementation of a low-power and efficient Smart Work Zone network using BLE in the context of emerging smart cities and advanced Internet of Things applications in transportation systems. Smart Work Zones significantly enhance road safety, improve efficiency, and support environmental sustainability. To this extent, sensing and communicating nodes are fitted to the work zone barrels to provide real-time information. However, work zone connectivity is a challenging task, as the implemented solution needs to be low-cost, low-maintenance, and requires low energy. 

In this study, a novel relay node selection algorithm is proposed for smart work zone barrels connectivity. The algorithm builds on the basic BLE mesh specification by introducing a relay node selection approach. The proposed algorithm operates by forming clusters of nodes, each of which is assigned a relay node responsible for receiving data transmissions from the other nodes within the cluster. 
 In collaboration with the Florida Department of Transportation (FDOT), this research was guided by three key constraints: low cost, low power consumption, and minimal maintenance. The low-cost requirement was critical due to the possibly high number of barrels that would need to be equipped with sensors. Additionally, since it would be impractical to frequently replace or recharge the batteries of these sensors, the design required the use of low-power strategies to extend battery life and minimize maintenance efforts. Another significant consideration in our network design is the semi-static nature of work zones. Unlike completely ad hoc networks, such as mobile or vehicular networks, or fully static networks, like those used in bridges or pipelines, work zone networks are characterized by their temporary and potentially variable configurations. Factors such as the relocation of barrels or external impacts (e.g., vehicles colliding with barrels) can change the layout and length of the work zone. Hence, the network needs to efficiently broadcast real-time information, such as the location of each barrel. However, continuous broadcasting would negatively impact the low-energy objective. To address this, our approach takes into consideration that nodes might only transmit data when necessary, such as when a barrel is moved from its original position, which helps conserve energy. On the other hand, for this strategy to work, the network must guarantee that once a node transmits, its data is successfully received by the central node. Therefore, our design enhances the message delivery ratio across all nodes, regardless of their distance from the central node, while simultaneously optimizing power consumption.

 The results of the study demonstrate that the proposed algorithm significantly outperforms three other relay node selection approaches, with packet delivery ratios reaching above 80\% in certain scenarios while maintaining low power consumption. Furthermore, the proposed algorithm shows an improvement in the per-node packet delivery ratio, which ensures that the destination is reachable by all nodes, especially those outside its communication range. The proposed algorithm also shows improvement in the relay load distribution amongst the relay nodes compared to other approaches.
Overall, the proposed relay node selection algorithm represents a promising approach to creating reliable and efficient data transmission in smart work zones. 

This work has multiple practical implications such as enhanced connectivity in smart work zones, as the proposed algorithm's enhanced data communication reliability directly translates to improved work zone efficiency and safety. Reliable data transmission ensures that safety alerts are promptly received by decision-makers. Hence, reducing the likelihood of hazardous situations and enhancing the overall safety of both workers and commuters within a work zone environment.
Another practical application of this work is efficient work zone management, where work zone managers can better monitor and control the work zone with real-time and reliable data communication. It opens the door to integrating various IoT sensors in work zone equipment for data collection.
This leads to a more efficient use of resources and minimizes potential traffic disruptions.

In conclusion, the proposed relay node selection algorithm plays a pivotal role in advancing smart work zone development. Enhancing the reliability and efficiency of data transmission lays the groundwork for creating smart work zones that are not only safer but also more efficient. This improvement ensures that decision-makers receive critical data promptly, facilitating quicker and more informed decisions for work zone safety management.

\bibliographystyle{IEEEtran}
\bibliography{ref}

\newpage

 




\begin{IEEEbiography}[{\includegraphics[width=1in,height=1.25in,clip,keepaspectratio]{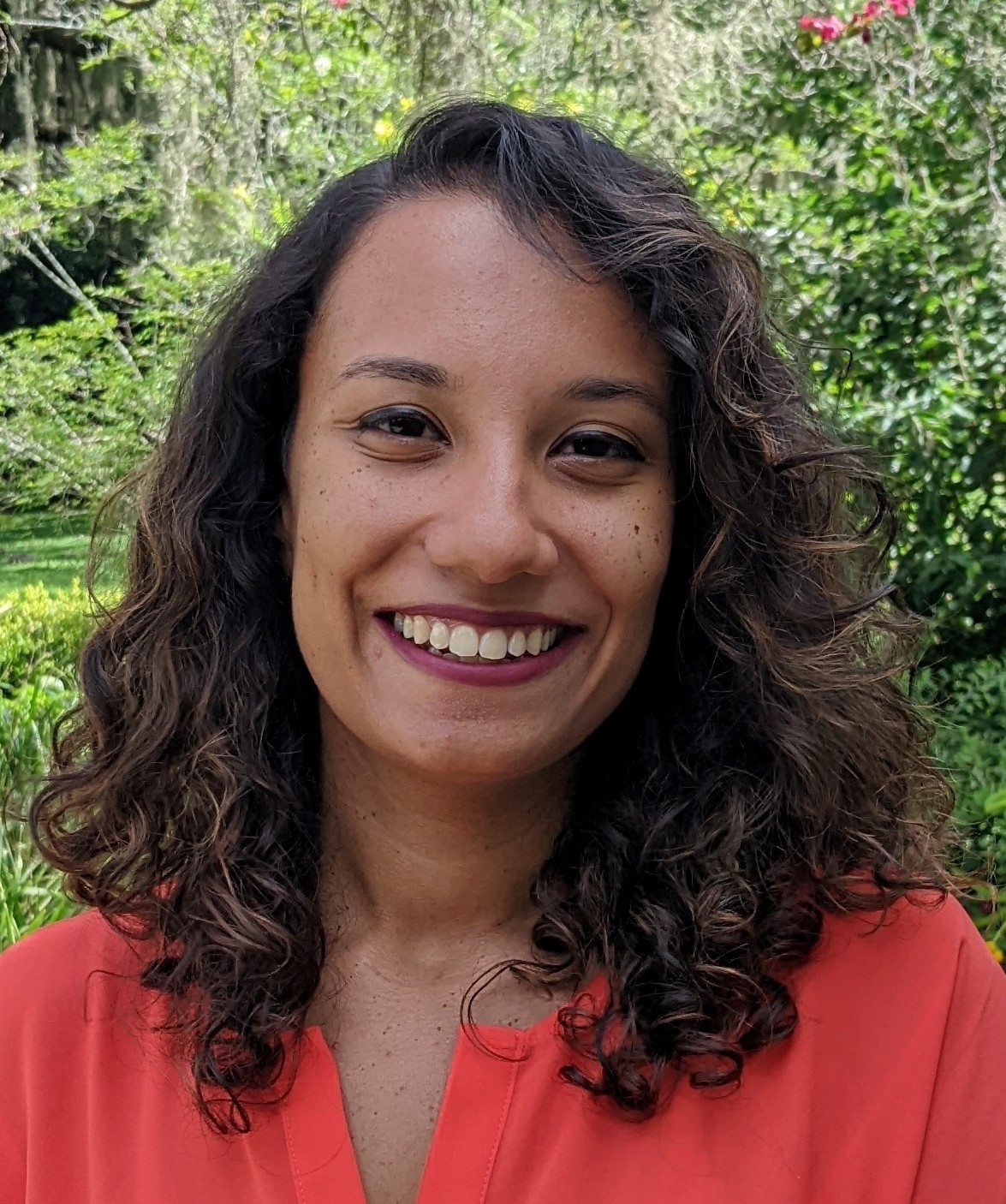}}]{Mariam Nour} is a PhD candidate in the Department of Civil, Environmental and Construction Engineering at the University of Central Florida. She has received her B.Sc degree in computer network engineering from the German University in Cairo (2017) and her M.Sc. degree in Smart Cities from the University of Central Florida (2021). 
In summer 2020, she was a Data Science intern in Connected Wise, R\&D Department in Orlando, FL. During the summer of 2024, she was an intern in Toyota InfoTech Labs in Mountain View, CA. Her research interests include road safety, vehicular connectivity and wireless sensor networks.
Mrs. Nour is the co-founder and vice-president of the Women's Transportation Seminar UCF Student Chapter (WTS-UCF). She was the recipient of the UCF ORC Fellowship in 2021 and the WTS Central Florida Chapter's Frankee Hellinger Graduate Scholarship in 2022.

\end{IEEEbiography}

\begin{IEEEbiography}[{\includegraphics[width=1in,height=1.25in,clip,keepaspectratio]{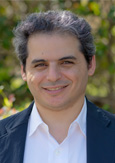}}]{Mohamed H Zaki} (Member, IEEE) is an Assistant Professor in the Civil and Environmental  Engineering Department at Western University in Ontario, Canada. Before, he was an assistant professor in Civil, Environmental \& Construction Engineering Department at the University of Central Florida, Orlando, Florida. Dr. Zaki was a research associate at the Bureau of Intelligent Transportation Systems and Freight Security at the University of British Columbia. He received his Doctoral degrees at the Hardware Verification Group from Concordia University, Montreal, in 2008. Dr. Zaki multidisciplinary research focuses on solving tomorrow's smart cities problems; from the computing and information to its facilities infrastructure. He studies road safety and road-users' behavior through the automated analysis of traffic data. Dr. Zaki is an IEEE member and serves on the Transportation Research Board (TRB) AED50 Committee on Artificial Intelligence and Advanced Computing Applications.
\end{IEEEbiography}

\begin{IEEEbiography}
[{\includegraphics[width=1in,height=1.25in,clip,keepaspectratio]{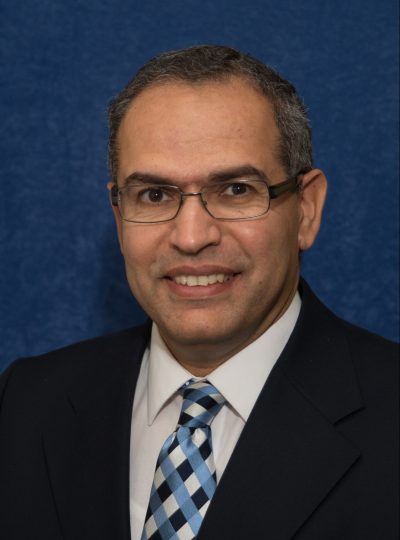}}]
{Mohamed Abdel-Aty} (Senior Member, IEEE) is
currently a Trustee Endowed Chair with the University of Central Florida (UCF). He is a Pegasus Professor and the Chair of the Civil, Environmental and Construction Engineering Department, UCF, where he is leading the Future City Initiative. He is also the Director of Smart and Safe Transportation Lab. He has authored or coauthored more than 825 papers, more than 430 in journals. As of Jan 2024, his Google Scholar citations are 31,400 and H-index is 98. His main expertise and interests are in the areas of traffic safety, simulation, big data and data analytics, ITS, and CAV. He was the recipient of 12 Best Paper awards.

\end{IEEEbiography}



\vfill

\end{document}